\documentclass[11pt]{amsart}
\usepackage[margin=1.05in]{geometry}
\usepackage{enumerate}
\usepackage[dvips]{epsfig}
\usepackage{amssymb}
\usepackage{graphicx}

\numberwithin{equation}{section}
\theoremstyle{plain}
\newtheorem{theorem}{Theorem}%[section]
\newtheorem{lemma}[theorem]{Lemma}

\theoremstyle{definition}

\newtheorem{conj}[theorem]{Conjecture}

\theoremstyle{remark}

\newcommand{\suppress}[1]{}
\newcommand{\zset}{\mathbb Z}
\newcommand{\fset}{\mathbb F}

\newcommand{\cset}{\mathbb C}

\newcommand{\balpha}{{\alpha^*}}
\numberwithin{equation}{section}
\newcommand{\ep}{{\varepsilon}}

\newcommand{\be}{\begin{equation}}
\newcommand{\ee}{\end{equation}}
\newcommand{\bea}{\begin{eqnarray}}
\newcommand{\eea}{\end{eqnarray}}
\newcommand{\bean}{\begin{eqnarray*}}
\newcommand{\eean}{\end{eqnarray*}}
\newcommand{\da}{\dagger}

\renewcommand{\Re}{{\mathfrak R}}
\newcommand{\abs}[1]{\left| #1 \right|}

\usepackage{color}

\title{Tree Codes and a Conjecture on Exponential Sums}

\author{Cristopher Moore}
\author{Leonard J. Schulman}
\date{\today .
 \emph{Author affiliations:} Santa Fe Institute, \tt{moore@santafe.edu}; Caltech, \tt{schulman@caltech.edu}}
\begin{document}
\pagenumbering{gobble}
\clearpage
\thispagestyle{empty}

\begin{abstract}
We propose a new conjecture on some exponential sums. These particular sums have not apparently been considered in the literature. Subject to the conjecture we obtain the first effective construction of asymptotically good tree codes. The available numerical evidence is consistent with the conjecture and is sufficient to certify codes for significant-length communications.
\end{abstract}

\maketitle
\clearpage
\pagenumbering{arabic}

\section{Introduction}
\label{sec:introduction}
Tree codes are a class of error-correcting codes that were introduced in~\cite{Schulman93STOC,coding-inter-comm} as the main combinatorial tool underlying the interactive coding theorem. They play a role in general (two-way) communication protocols that is analogous to the role of block error correcting codes in Shannon's theory of (one-way) message transmission.

In spite of the existence proof provided in~\cite{Schulman93STOC,coding-inter-comm}, and in spite of efforts by researchers in the interim, to date no effective construction of tree codes is known. By ``effective construction'' we mean a deterministic algorithm that provides any requested code-word entry within time polynomial in the address-length of that entry (see detailed definitions below). In fact, even randomized constructions are unknown. The best known results have been, first, elementary ideas giving polynomial (rather than constant) size alphabet or else providing a weaker tree code property~\cite{post-coding-inter-comm}; a modified existence proof with slightly better parameters by Peczarski~\cite{DBLP:journals/ipl/Peczarski06}; and a construction of Braverman~\cite{Braverman12} giving a tree of depth $n$ in time $2^{n^\ep}$.

Because of this gap, researchers have come up with several innovative ways to skirt the challenge of constructing tree codes yet still obtain noise robustness for communication protocols. Ostrovsky, Rabani, and Schulman~\cite{OstrovskyRS09} designed effective codes that can replace tree codes in a restricted class of communication protocols.
Gelles, Moitra, and Sahai~\cite{GellesMS11} introduced another class of codes, also weaker than tree codes, called ``potent tree codes'', which are again sufficient to recover the coding theorem; they showed that a random construction suffices to provide a potent tree code with high probability, but deterministic constructions remain unknown even for this weaker object.

In spite of these considerable advances in protocol design, the basic problem of effectively and deterministically constructing the underlying class of
tree codes has remained open since 1993. This question will still remain open after the current paper, but we present here what is the first plausible candidate for an effective (i.e., deterministic, polynomial-time) construction of tree codes. The validity of the construction, namely whether the codes actually satisfy the required minimum-distance condition, depends on a conjecture about some particular exponential sums. Apparently, no bounds for these sums are known. However, analogous sums have drawn considerable attention, which we will discuss in Sec.~\ref{sec:block}.

Outline of the paper: In Sec.~\ref{sec:prelim} we recall the definition of tree codes.
In Sec.~\ref{sec:block}, as a warm-up, we show an elementary application of our method to the classic problem of constructing block codes. This will demonstrate that an extremely simple construction of asymptotically good block codes over bounded alphabets would be implied by an correspondingly bold conjecture about short exponential sums (of the type that has already been studied in analytic number theory). This section has no direct bearing on tree codes, however.
In Sec.~\ref{sec:construction} we provide our candidate construction of tree codes and our associated conjecture. In Sec.~\ref{sec:connex} we disgress to discuss the relationship between our construction and the long-studied question of the distribution of the sequence $\xi \theta^n \bmod 1$, in particular for $\theta=3/2$. 
In Sec.~\ref{sec:numerics} we present numerical tests of our conjecture that rule out small counterexamples. In particular our own tests (which required only a few days of machine time) validate the use of our codes for communication protocols up to $90$ rounds, well beyond the range at which the algorithms for other constructions can be validated with comparable computational effort, and similar calculations by students went even further and validate our codes up to $145$ rounds.

\subsection{Other literature}
A few words on how tree codes are used. In~\cite{Schulman93STOC,coding-inter-comm}, tree codes were introduced as part of a particular protocol to simulate noiseless channels on noisy ones. This protocol is computationally efficient in the case of stochastic noise; the protocol also works for adversarial noise, up to fractional Hamming distance $1/240$, but in this case is not computationally efficient. Braverman and Rao~\cite{BravermanRao11} improved the distance bound to $1/4$ by modifying how the protocol uses the underlying tree codes. 
In~\cite{FranklinGOS13} tree codes were used both for cryptographic ends, namely to authenticate data streams and interactive protocols, and in order to improve the above $1/4$ to $1/2$ under the additional assumption of shared randomness. 
Brakerski and Kalai~\cite{BrakerskiK12} and Brakerski and Naor~\cite{BrakerskiN13} showed how to handle even adversarial error with low computational overhead (first polynomial and then near-linear). In these recent two papers, tree codes of logarithmic depth (in the number of rounds of communication) are all that is needed, so (depending on the noise level and the constants in the protocols), our numerical results may be enough to enable some applications. 

\section{Preliminaries}
\label{sec:prelim}

The infinite binary rooted tree $T$ is the undirected graph whose vertices are the finite binary sequences $\{0,1\}^*$, and in which each vertex $x=(x_1,\ldots,x_n) \in \{0,1\}^n$ is connected to its parent $(x_1,\ldots,x_{n-1})$.  We say $x$ is at level $n$, and we let $L_n = \{0,1\}^n$ denote the set of vertices at level $n$. The root is the empty sequence $\ep$, at level $0$.

We are interested in $T$ as a \emph{leveled} metric space, that is to say, we define a distance function on each set $L_n$ but use a common notation $d_T$.
If $x, y \in L_n$ and
$\ell + 1$ is the least integer $i$ such that $x_i \neq y_i$, then we set $d_T(x,y)=n-\ell$.
(This metric can be straightforwardly
extended to a distance function on the full tree but we shall not need this.)

Let $\Sigma$ be any alphabet. The set of finite sequences $\Sigma^*$ is also a leveled metric space, using Hamming distance on strings of the same length: if
$\sigma, \tau \in \Sigma^n$,
then $d_\Sigma(\sigma,\tau)=|\{i: \sigma_i \neq \tau_i \}|$.

Let $\alpha$ be a mapping from $\bigcup_{n>0} L_n$ to $\Sigma$, assigning a label in $\Sigma$ to every node in $T$ other than the root.  This induces a secondary mapping $\balpha$ from the vertices of $T$ into $\Sigma^*$, where $\balpha(x)$ is the sequence of labels on the path from the root to $x$.  That is, $\balpha(\ep) = \ep$, and for $n > 0$, $x \in L_n$,
\[  \balpha(x_1,\ldots,x_n)=\balpha(x_1,\ldots,x_{n-1}) \circ \alpha(x_1,\ldots,x_n) \, , \]
where $\circ$ denotes composition.  In particular, $\balpha$ maps $L_n$ into $\Sigma^n$.

We say that $\alpha$ satisfies the \emph{tree code distance property} if $\balpha$ is bi-Lipschitz: that is, if there is a constant $c > 0$ such that for all $n \ge 0$ and all $x,y \in L_n$,
\[ d_\Sigma(\balpha(x),\balpha(y)) \geq c d_T(x,y) \, . \]
There is no particular need for the host metric $d_\Sigma$ to be the Hamming distance, but for bounded alphabets the choice matters little.  Below we will embed sequences $\Sigma^*$ in the complex numbers $\cset$, in which case we will replace $d_\Sigma$ with the usual norm on $\cset$.

We say that $\alpha$ is a \emph{deterministic, effective, asymptotically good tree code} if $|\Sigma| < \infty$, the tree code distance property is satisfied,
and $\alpha$ is computable in deterministic polynomial time.  The question of whether such codes exist (even if we replace deterministic with randomized computation) is still open.  Our goal in this paper is to present a construction $\alpha$, and a number-theoretic conjecture regarding exponential sums, such that $\alpha$ is indeed asymptotically good if the conjecture is true.

\section{An illustration: block codes}
\label{sec:block}

As a warm-up, in this section we show how the simple step of embedding the vertices of $T$ in the complex numbers $\cset$ can be applied toward a much simpler problem, namely the construction of asymptotically good block error codes---assuming that a certain conjecture regarding short exponential sums is true.

A probabilistic method argument due to Shannon in 1948~\cite{Shannon48aMTC} shows the existence of asymptotically good block codes---namely, mappings from $\{0,1\}^n$ to $\Sigma^n$ for $\Sigma$ independent of $n$, such that for some $c > 0$ every two points in the image of the map have Hamming distance at least $cn$.  It was not until 1966 that Forney~\cite{Forney66} (improved by Justesen in 1972~\cite{Justesen72}) gave a construction for effectively computable codes with bounded alphabet size, this by first making the construction in a finite field of unbounded size (roughly $n$), and then concatenating with a secondary code that maps each element of the finite field to a string over a bounded-size alphabet.  (In~\cite{Justesen72}, this code is location-dependent.)

The construction we describe now is simpler, although somewhat analogous: again we use as intermediary a code over a large field, in this case $\cset$. The secondary mapping now carries each field element to a single character (rather than a sequence) in a bounded-size alphabet $\Sigma=\{0,\ldots,\kappa-1\}$. Of course this secondary mapping can no longer be injective.

We treat sequences $\{0,1\}^n$ as elements of the cyclic group $\zset/2^n$.  Then the construction is a composite mapping
\[ \gamma^{n'} \circ (\beta_1,\ldots,\beta_{n'}): \zset/2^n \to \Sigma^{n'} \, , \]
for $n'=cn$ where $c$ is a sufficiently large constant.  Here each $\beta_k$ maps $\zset/2^n$ into
$\cset$,
$\gamma$ maps the unit circle in $\cset$ to $\Sigma$, and
$ \gamma^{n'}$ is the $n'$-fold direct product of $\gamma$ with itself. Specifically, let
\[
e(x) = \exp(2 \pi i x) \, .
\]
Then $\beta_k$ is defined by
\[
\beta_k(m) = e\!\left( 3^k \frac{m}{2^n} \right)
\]
for each $0 \le k < cn$.  Observe that for any $m, m' \in \zset/2^n$,
\[
\beta_k(m) \,\beta_k(m')^\dagger = \beta_k(m-m')
\]
where $\dagger$ denotes complex conjugation.  
Finally, recalling that $\Sigma = \{0,\ldots,\kappa-1\}$, $\gamma$ rounds each component $z_k$ according to which of the $\kappa$ intervals of width $2\pi/\kappa$ it lies in:
\[
\gamma(z_k) = \ell
\quad \text{such that} \quad
-\pi/\kappa \le \arg \frac{z_k}{e(\ell/\kappa)} < \pi/\kappa \, .
\]

Below we give a linear lower bound on the Hamming distance $d_\Sigma$ between $(\gamma^{cn} \circ (\beta_1,\ldots,\beta_{cn}))(m)$ and $(\gamma^{cn} \circ (\beta_1,\ldots,\beta_{cn}))(m')$ for $m, m' \in \zset/2^n$ conditional on the following conjecture.
\begin{conj}
\label{conj:block}
There are constants $c < \infty$ and $\delta > 0$ such that for all $n$ and all nonzero $m \in \zset/2^n$,
\begin{equation}
\label{eq:conj1}
\Re \left( \frac{1}{cn} \sum_{0 \leq k < cn} e\!\left( 3^k \frac{m}{2^n} \right) \right) \leq 1-\delta \, .
\end{equation}
\end{conj}
\noindent
(where $\Re$ denotes real part).  We emphasize that we are unaware of implications in either direction between this conjecture and Conjecture~\ref{conj:main-conj} (or \ref{conj:main-conj-z}), which are the main focus of this paper.  

We make several remarks about the exponential sum appearing in this conjecture.  First, note that it is the Fourier transform, at frequency $m \in \zset/2^n$, of the uniform distribution on the geometric series $\{1,3,3^2,\ldots,3^{cn-1}\}$.  
If we replace the modulus $2^n$ with a prime $p$ so that $\zset/2^n$ becomes the finite field $\fset_p$, and replace $3$ with a primitive root $g$, 
then results of Korobov~\cite{korobov} give the following bound on the absolute value (which is obviously also a bound on the real part):
\[
\abs{ \frac{1}{N} \sum_{0 \le k < N} e( g^k m/p ) } \le \frac{p^{1/2} \log p}{N} \, .
\]
However, this bound becomes nontrivial only when $N > p^{1/2} \log p$.
Improved bounds were given by Bourgain and Garaev~\cite{bourgain-garaev} and Konyagin and Shparlinski~\cite{konyagin-shparlinski}; see also Bourgain and Glibichuk~\cite{BourgainG11} for complete sums over multiplicative subgroups, i.e., over all the powers of some element $g$ that is not a primitive root, and Kerr~\cite{kerr} for incomplete sums of this type.  However, all of these only become nontrivial when $N \ge p^\alpha$ for some $\alpha > 0$.

In contrast, we are interested in the average of the first $N=cn$ terms of the sum (for any sufficiently large constant $c$), which corresponds to $O(\log p)$ terms in the finite field case; in this regime, despite the results cited and related recent advances in additive combinatorics, nothing appears to be known. 

Note that, at the least, Conjecture~\ref{conj:block} requires that $c>\log_3 2$. Otherwise, taking $m=1$, most terms of the summation are close to $1$, as is the real part of their average. For the very modest value $n=20$ and for $c=2$ we have computed the LHS of Eqn.~\ref{eq:conj1} for all $m$, and the evidence is consistent with Conjecture~\ref{conj:block}, and even makes plausible that the distribution of the LHS (with $m$ uniformly random) tends to normal, with variance sufficiently small for the minimum distance guarantee. This is a phenomenon familiar in coding theory (see ~\cite{MacWilliamsS77} p.~287 or \cite{KrasikovL98} and references therein).

As another remark, note that the geometric series appearing in Eqn.~\ref{eq:conj1} is generated by the fixed element $3$, no matter the value of $n$.  Arbitrary geometric series will not do.  Consider replacing the generator $3$ by $3^{2^r}$. It is known that the order of $3$ modulo $2^n$ is $2^{n-2}$ for $n\geq 3$. As a consequence, for $r \geq 1$, $3^{2^r}$ generates a multiplicative cyclic group $G \subset \zset/2^n$ which equals the additive coset $1+2^{r+2}\zset$.  Now consider the additive subgroup $H=2^{n-r-2}\zset$.
If $m \notin H$, the sum in Eqn.~\ref{eq:conj1} is zero; however, if $m \in H$, it is $e(m/2^n)$.  As we vary $m$ in the latter case, these values are spread evenly on the unit circle and in particular there are (many) nonzero $m$ with real part arbitrarily close to $1$.

Despite having little confidence in, and very limited evidence for, Conjecture~\ref{conj:block}, we plow ahead, showing that, if true, it implies that the above construction is an asymptotically good block error-correcting code. The following lemma (which we later employ again in the tree code section) provides the connection with Hamming distance.

\begin{lemma}
\label{lem:inprod-to-hamming}
Let $z=(z_1,\ldots,z_{n'})$ and $z'=(z'_1,\ldots,z'_{n'})$ with
$|z_k| = |z'_k| = 1$
for all $1 \le k \le n'$. Then
\[ d_\Sigma\big( \gamma^{n'}(z), \gamma^{n'}(z') \big)
\geq \eta n'
\quad \text{where} \quad
\eta = \frac {\cos 2 \pi / \kappa - (1/n') \Re \sum_k z_k {z'_k}^\da  }
{1+ \cos 2 \pi / \kappa} \, .
\]
\end{lemma}

\noindent
(Note this bound is meaningful only for $\kappa > 4$.)

\begin{proof}
Observe that if $\gamma(z_k)=\gamma(z'_k)$ then
$\Re\left( z_k {z'}^\da_k \right) \geq 1-\cos 2 \pi / \kappa$.
On the other hand for any
$z_k$, $z'_k$ we have
$\Re\left( z_k{z'_k}^\da \right) \geq -1$.
Therefore, with $h=d_\Sigma\big( \gamma^{n'}(z), \gamma^{n'}(z') \big)$,
\begin{align*}
\frac{1}{n'} \,\Re \sum_{k=1}^{n'} (1- z_k {z'}^\da_k)
&= 1 - \frac{1}{n'} \sum_k \Re \left( z_k {z'}^\da_k \right) \\
&\leq 1-\cos 2 \pi / \kappa + \frac{h}{n'} (1+ \cos 2 \pi / \kappa) \, ,
\end{align*}
and rearranging completes the proof.
\end{proof}

Finally, if Conjecture~\ref{conj:block} is true with a given $c$ and $\delta$, we choose $\kappa$ such that $\cos 2 \pi / \kappa > 1-\delta$.  Then Lemma~\ref{lem:inprod-to-hamming} tells us that $(\gamma^{n'} \circ (\beta_1,\ldots,\beta_{n'}))(\zset/2^n)$ is an error-correcting code of length $n'=cn$ and distance $\eta n'$ where $\eta > 0$.

\section{The tree code construction and the main conjecture}
\label{sec:construction}

Even if the rather strong conjecture in the preceding section can be proven, coding theory has of course long had other effective constructions of block codes. The main purpose of that section was to prepare for the following construction of tree codes.

Define a mapping $\beta$ from $\{0,1\}^*$ to $\cset$ as follows.  Set $\beta(\ep) = 1$.  If $k>0$, $\beta$ has been defined on $L_{k-1}$, and $x=(x_1,\ldots,x_k) \in L_k$, define $\beta(x)$ inductively using
\[
\beta(x_1,\ldots,x_{k})^2=
\beta(x_1,\ldots,x_{k-1})^3 \, ;
\]
This equation has of course two roots, and we use one for each child of $(x_1,\ldots,x_{k-1})$.  Specifically, let $b \in [0,1)$ such that
\[
\beta(x_1,\ldots,x_{k-1}) = e(b) \, ,
\]
i.e., $b = (1/2\pi) \arg \beta(x_1,\ldots,x_{k-1})$.  Then make the assignment to each of the children following the arbitrary convention
\begin{equation}
\label{eq:convention}
\beta(x_1,\ldots,x_k) = e\!\left( \frac{3}{2} b
+ \frac{x_k}{2} \right) \, .
\end{equation}
Now let $\alpha=\gamma \circ \beta$, with $\gamma:\cset \to \Sigma$ the rounding function defined in the previous section, and define $\balpha: T \to \Sigma^*$ as in Sec.~\ref{sec:prelim}.

A key feature of this tree code design is that, in analogy with (one of the) original existence proofs in~\cite{Schulman93STOC,coding-inter-comm}, it has the property that its metric properties can be understood solely
in terms of
what happens to pairs of paths that diverge at the root. 
The convolutional construction in~\cite{Schulman93STOC,coding-inter-comm}, let us call it $\tau$, had the following ``translation invariant'' property: 
Suppose that $x,x' \in L_n$ and that $d_T(x,x') = \ell$. Write
\begin{align} \label{x,x'}
x &= (x_1,\ldots,x_{n-\ell},1,x_{n-\ell+2},\ldots,x_n) \\
x' &= (x_1,\ldots,x_{n-\ell},0,x'_{n-\ell+2},\ldots,x'_n) \, . \nonumber
\end{align}
Then 
\[
d_\Sigma\big( \tau^*(x),\tau^*(x') \big) = d_\Sigma\big( \tau^*(1,x_{n-\ell+2},\ldots,x_n),\tau^*(0,x'_{n-\ell+2},\ldots,x'_n) \big) \,. 
\]
That is to say, Hamming distances in $\tau$ are ``translation invariant'' as we shift pairs of paths about the tree: adding, removing, or changing their common initial subsequence.  In particular, they are invariant if we remove their common initial subsequence completely so they diverge at the root.

This translation invariance \textit{almost} holds in the present construction. Specifically,
the inner product employed in Lemma~\ref{lem:inprod-to-hamming} to lower bound the Hamming distance, does have this translation invariance. The invariance is lost only in the rounding by $\gamma$. Therefore it suffices to study inner products for pairs of paths which diverge at the root. The construction also has additional symmetry which we discuss below.

The lower bound method for our construction is as follows. Let $x,x' \in L_n$ be as in Eqn.~(\ref{x,x'}).
Then
\[
d_\Sigma\big( \balpha(x),\balpha(x') \big)
= \sum_{k=n-\ell+1}^n d_\Sigma(\gamma \beta(x_1,\ldots,x_k),\gamma \beta(x'_1,\ldots,x'_k)) \, .
\]
By Lemma~\ref{lem:inprod-to-hamming} (with $\ell$ here corresponding to $n'$ in the lemma), this is bounded below by
\begin{equation}
d_\Sigma\big( \balpha(x),\balpha(x') \big)
\geq
  \frac {\cos 2 \pi / \kappa - (1/\ell) \Re \sum_{k=n-\ell+1}^n
  \beta(x_1,\ldots,x_k) \,\beta(x'_1,\ldots,x'_k)^\da}
{1+ \cos 2 \pi / \kappa} \, \ell \, .
\label{eq:kappa}
\end{equation}
We can now eliminate the common initial subsequence in $x_1,\ldots,x_{n-\ell}$ from this bound; this lets us reduce to the case where the paths diverge at the root.  Let $b \in [0,1)$ such that
\[
\beta(x_1,\ldots,x_{n-\ell}) = e(b) \, .
\]
Then~\eqref{eq:convention} gives
\begin{align*} \sum_{k=n-\ell+1}^n & \beta(x_1,\ldots,x_k) \,\beta(x'_1,\ldots,x'_k)^\da \\
= \;& e\!\left(\frac{3}{2}b+\frac{1}{2} \right) \,e\!\left(-\frac{3}{2}b \right) \\
& + e\!\left(\frac{3}{2}\left( \frac{3}{2}b+\frac{1}{2}\right)+\frac{x_{n-\ell+2}}{2} \right)
\,e\!\left(-\frac{3}{2}\left( \frac{3}{2}b \right)-\frac{x'_{n-\ell+2}}{2} \right) \\
& +  e\!\left(\frac{3}{2}\left( \frac{3}{2}\left( \frac{3}{2}b+\frac{1}{2} \right)+\frac{x_{n-\ell+2}}{2} \right)+\frac{x_{n-\ell+3}}{2} \right)
e\!\left(-\frac{3}{2}\left( \frac{3}{2} \left( \frac{3}{2}b \right)+\frac{x'_{n-\ell+2}}{2} \right)-\frac{x'_{n-\ell+3}}{2} \right) + \cdots \\
= \;& e\!\left( \frac{1}{2} \right)
+ e\!\left( \frac{3}{2} \frac{1}{2}+\frac{x_{n-\ell+2}-x'_{n-\ell+2}}{2} \right) \\
& + e\!\left(\left( \frac{3}{2} \right)^{\!2} \frac{1}{2}+ \left( \frac{3}{2} \right) \frac{x_{n-\ell+2}-x'_{n-\ell+2}}{2} + \frac{x_{n-\ell+3}-x'_{n-\ell+3}}{2} \right) + \cdots \\
 & + e\!\left(\left( \frac{3}{2} \right)^{\!\ell-1} \frac{1}{2}+\left( \frac{3}{2} \right)^{\!\ell-2} \frac{x_{n-\ell+2}-x'_{n-\ell+2}}{2}
 + \left( \frac{3}{2} \right)^{\!\ell-3} \frac{x_{n-\ell+3}-x'_{n-\ell+3}}{2} + \cdots +\frac{x_{n}-x'_{n}}{2}
 \right) \\
= \;& \sum_{i=1}^{\ell} e\!\left( \frac{1}{2} \left( \left( \frac{3}{2} \right)^{\!i-1} + \sum_{j=2}^i \left( \frac{3}{2} \right)^{\!i-j} (x_{n-\ell+j}-x'_{n-\ell+j}) \right) \right)
\, .
\end{align*}

Now write $y_j = x_{n-\ell+j}-x'_{n-\ell+j}$.  Then in order to show that $\alpha$ is a deterministic, effective, asymptotically good tree code, it is sufficient to show the following:
\begin{conj}
\label{conj:main-conj}
There is a $\delta>0$ such that for any $\ell \geq 1$ and any sequence $y_1,\ldots,y_\ell$ with   $y_1=1$ and all $y_j \in \{-1,0,1\}$,
\begin{equation}
\Re \left[ \frac{1}{\ell} \sum_{i=1}^\ell e\!\left( \frac{1}{2} \sum_{j=1}^{i} \left( \frac{3}{2} \right)^{\!i-j} y_j \right) \right]
\leq 1-\delta \, .
\label{eq:main-conj}
\end{equation}
\end{conj}
Before proceeding further we point out a computational implication of the sufficiency of this conjecture. The fractional-distance bound for a tree code of depth $n$ is the minimum of $\Theta(4^n)$ pairwise distances; in general, for a proposed tree code, verification to depth $n$
may require this many calculations. But for the present construction, it is evidently sufficient to check $\Theta(3^n)$ distances. (Further improvement will be discussed below.)

It is easier to read Conjecture~\ref{conj:main-conj} as being about exponential sums if we define
$e_N : \zset/N \to \cset$ as $e_N(z)=e(z/N)$; note that all arithmetic in the argument of $e_N$ is performed in $\zset/N$. Now if (for $y_1,\ldots,y_\ell$ with   $y_1=1$) 
 we write
\begin{equation}
\label{eq:z}
z =
\sum_{j=1}^{\ell} 2^{j-1} 3^{\ell-j} y_j \in \zset \, ,
\end{equation}
then we have
\[
e\!\left( \frac{1}{2} \sum_{j=1}^{i} \left( \frac{3}{2} \right)^{\!i-j} y_j \right)
= e_{2^i}\!\left( \sum_{j=1}^{i} 2^{j-1} 3^{i-j} y_j \right)
= e_{2^\ell}\!\left( \left(\frac{2}{3}\right)^{\!\ell-i} z \right)
\]
(where in the argument of $e_{2^\ell}$, negative powers of $3$ are of course positive powers of $3^{-1} \bmod 2^\ell$). 
Note that the terms $i < j \le \ell$ in the last sum disappear, as they are divisible by $2^\ell$.
Substituting $m=\ell-i$ transforms the left-hand side of~\eqref{eq:main-conj} into
\[
\Re \left[ \frac{1}{\ell} \sum_{m=0}^{\ell-1} e_{2^\ell} \!\left( \left(\frac{2}{3}\right)^{\!m} z \right) \right] \, .
\]

Inspecting~\eqref{eq:z}, we see that our condition that $y_1 = 1$ forces $z$ to be odd; conversely, any odd $z \in \zset/2^\ell$ corresponds to a sequence $y_1=1, y_2, \ldots, y_\ell$ (and usually to many such sequences).
This is easy to see by induction. Let $z' \in \zset/2^{\ell-1}$ satisfy $3z'=z \!\pmod{2^{\ell-1}}$ and let 
$y_1, y_2, \ldots, y_{\ell-1}$ be a sequence to which $z'$ corresponds as in~\eqref{eq:z}. Then an appropriate choice of $y_\ell$ extends the representation of $z'$ to one of $z$.
This gives us the following equivalent form of Conjecture~\ref{conj:main-conj}:
\begin{conj}
\label{conj:main-conj-z}
There is a $\delta > 0$ such that for any $\ell \geq 1$ and any odd integer $z$,
\begin{equation}
\label{eq:main-conj-z}
\Re \left[ \frac{1}{\ell} \sum_{m=0}^{\ell-1} e_{2^\ell} \!\left( \left(\frac{2}{3}\right)^{\!m} z \right) \right]
\le 1-\delta \, .
\end{equation}
\end{conj}
Conjecture~\ref{eq:main-conj-z} has the following advantage over its equivalent sibling: verification of the conjecture for all $\ell \leq n$ (which is all that is required in order to certify a tree code of depth $n$) requires only $\Theta(2^n)$ calculations, i.e., one for each $z$ between $1$ and $2^\ell$, rather than $\Theta(3^n)$.

We end this section with some remarks on the sum in~\eqref{eq:main-conj-z}.  We begin with an odd $z \in \zset/2^\ell$.  We repeatedly multiply it by $2 \cdot 3^{-1}$, producing a sequence $z_0 = z, z_1 = (2/3)z, z_2 = (2/3)^2 z, \ldots$\  Each such step maps $z$ into a smaller subgroup: $z_1$ is even, $z_2$ is a multiple of $4$, and so on until $z_{\ell-1} = 2^{\ell-1}$.  The function $e_{2^\ell}$ maps these onto the unit circle with arguments $2 \pi z/2^\ell$, and after $\ell-1$ steps we arrive at $\beta = e_{2^\ell}(2^{\ell-1}) = -1$.

Alternatively, we can view this process in reverse.  We start with $\beta = -1$, and then repeatedly map $\beta$ to one of the square roots of $\beta^3$, mapping its argument $\theta = \arg \beta$ to $(3/2) \theta$ or $(3/2) \theta+\pi$.  After $k$ steps, $\theta$ is a multiple of $2\pi/2^k$.  Taking $\ell-1$ steps produces $2^{\ell-1}$ different sequences of points on the unit circle, which correspond to the $2^{\ell-1}$ odd elements of $\zset/2^\ell$.

Now imagine an adversary seeking to maximize the real part of the sum in~\eqref{eq:main-conj-z}. In order to do this effectively he must create long runs of terms close to $1$ on the unit circle.  He can do this if he can generate a small argument $\theta$, so that successive arguments $(3/2) \theta, (3/2)^2 \theta, \ldots$ will also be small.  In particular, if on some step he generates $\theta = \Theta(2^{-k})$, then roughly the next $\log_{3/2} (1/\theta) = \Theta(k)$ terms will have arguments close to zero, contributing $1$ to the average.

At the end of such a run, the arguments will again be $\Theta(1)$, with large successive angles between terms.  However, if the adversary can generate an argument close to $2\pi/3$ or $4\pi/3$, i.e., a $\beta$ close to a cube root of $1$, he can generate another small argument on the next step, and start another run.  If he can do this repeatedly, so that a $1-o(1)$ fraction of the terms are close to $1$, then Conjecture~\ref{conj:main-conj-z} is false.  Thus our construction relies on the hope that he cannot design long sequences of runs with just a few terms between the end of each run and the start of the next.  Equivalently, our conjecture holds if there are constants $c, \phi > 0$ such that, for all $2^{\ell-1}$ sequences, at least $cn$ of the adjacent pairs of terms in~\eqref{eq:main-conj-z} have an angle $\phi$ between them.

\section{Related questions, and base-$3/2$ representations of one}
\label{sec:connex}

Since the early work of Pisot~\cite{Pisot38} and Vijayaraghavan~\cite{Vijayaraghavan40} (see~\cite{FlattoLP95} for more recent work and other references) it has been of interest in number theory to study the distribution of the sequence $\xi \theta^n \bmod 1$, and in particular for $\theta=3/2$. The extent of the relationship to our question is unclear, as we require an understanding of sequences generated by the operation $\times (3/2) + x/2$ for $x \in \{0,1\}$, whereas these works are restricted to $x=0$.

\begin{figure}
\begin{center}
\includegraphics[width=3in]{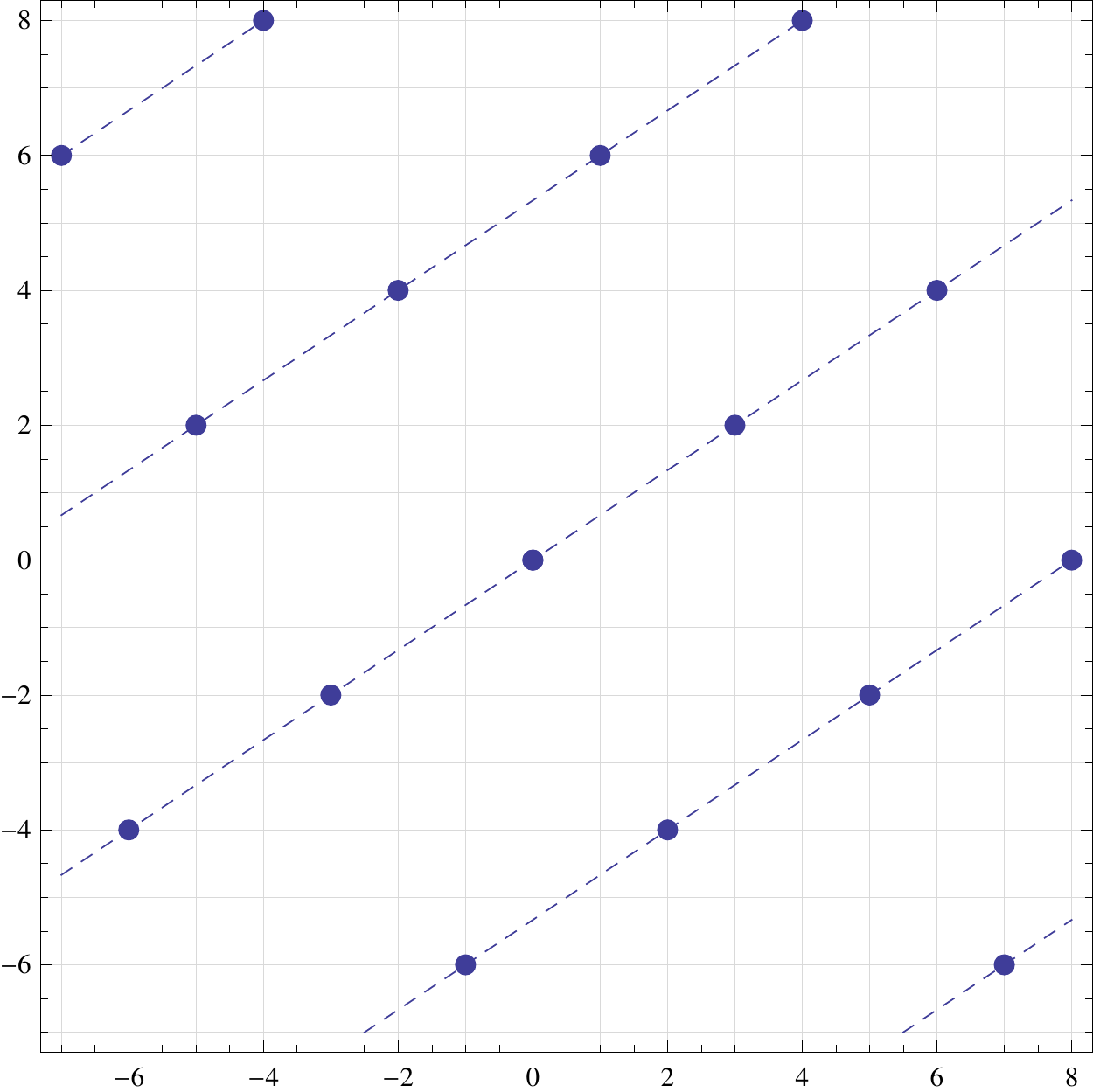}
\end{center}
\caption{The five branches of the function $(2/3)z$ when we identify $z \in \zset/2^\ell$ with an integer in the range $-2^{\ell-1} < z \le 2^{\ell-1}$.  Here $\ell=4$.}
\label{fig:twothirds}
\end{figure}

However, we can relate our main conjecture to a question of a similar flavor. Consider its statement in the form of Conjecture~\ref{conj:main-conj-z}.  Let us identify each element of $\zset/2^\ell$ with an integer in the range $-2^{\ell-1} < z \le 2^{\ell-1}$.  Then if
\[
z' \equiv_{2^\ell} \frac{2}{3} z \, ,
\]
(where as before $3^{-1}$ refers to the inverse of $3$ mod $2^\ell$) the corresponding integers obey
\[
z' = \frac{2}{3} \left( z + a 2^{\ell-1} \right)
\quad \text{for some} \quad
a \in \{-2,-1,0,1,2\} \, .
\]
In other words, the function $(2/3)z$ has several ``branches'', where the relevant branch depends on $z \bmod 3$, and also on whether $z$ is greater or smaller than $\pm 2^{\ell-2}$ due to wraparound effects; see Fig.~\ref{fig:twothirds}.

Now consider the sequence $z_0=z, z_1=(2/3)z, z_2 = (2/3)^2 z, \ldots$ discussed at the end of Section~\ref{sec:construction}.  We have
\begin{align*}
z_1 &= \frac{2}{3} \left( z + a_1 2^{\ell-1} \right) \\
z_2 &= \frac{2}{3} \left( \frac{2}{3} \left( z + a_1 2^{\ell-1} \right) + a_2 2^{\ell-1} \right) \\
&\;\;\vdots \\
z_{\ell-1} &= \left( \frac{2}{3} \right)^{\!\ell-1} z + 2^{\ell-1} \sum_{i=1}^{\ell-1} \left( \frac{2}{3} \right)^{\!\ell-i} a_i \, .
\end{align*}
However, we also know that $z_{\ell-1}=2^{\ell-1}$ since that is the only integer such that $-2^{\ell-1} < z \le 2^{\ell-1}$ and $z_{\ell-1} \equiv_{2^\ell} 2^{\ell-1}$.  Since $|z| \le 2^{\ell-1}$, rearranging then gives
\begin{equation}
\label{eq:two-thirds}
\abs{ 1 - \sum_{j=1}^{\ell-1} \left( \frac{2}{3} \right)^{\!j} a_{\ell-j} }
\le \left( \frac{2}{3} \right)^{\!\ell-1} \, .
\end{equation}
Thus the coefficients $a_i$ form a series that sums to $1$, with an error term that decreases as $(2/3)^\ell$.

There are certainly many such series, i.e., many ways to ``write $1$ in base $3/2$ with integer coefficients ranging from $-2$ to $2$".  However, we conjecture that, in any such series, a constant fraction of the coefficients $a_i$ must be nonzero: in other words, that the sequence is forced to jump from branch to branch of the function $(2/3)z$ a constant fraction of the time. Since the branches with $a \ne 0$ are all a constant fraction of $2^\ell$ away from $z$ in the range $-2^{\ell-1} < z \le 2^{\ell-1}$, this implies that a constant fraction of adjacent terms in~\eqref{eq:main-conj-z} have a constant angle between them.  Thus the following conjecture would imply Conjecture~\ref{conj:main-conj-z} (or equivalently Conjecture~\ref{conj:main-conj}):
\begin{conj}
\label{conj:two-thirds}
There is a constant $c > 0$ such that, if $\ell \ge 1$ and~\eqref{eq:two-thirds} holds where $a_i \in \{ -2, -1, 0, 1, 2 \}$ for all $i$, then $| \{ i : a_i \ne 0 \} | \ge c\ell$.  In particular, for any infinite sequence $a_1, a_2, \ldots$ with $a_i \in \{ -2, -1, 0, 1, 2 \}$ such that $\sum_{i=1}^\infty (2/3)^i a_i = 1$, any initial subsequence of length $\ell$ contains at least $c \ell$ nonzero terms.
\end{conj}

We offer no tools for proving Conjecture~\ref{conj:two-thirds}, but it seems of independent interest from a purely number-theoretic point of view. Perhaps it is too strong to be true; but if it fails while the main conjecture survives, that will be a fascinating distinction.  A distantly-related conjecture of Erd\H{o}s~\cite{erdos} states that there are only finitely many $n$ such that the ternary expansion of $2^n$ avoids the digit $2$.  Lagarias~\cite{Lagarias01062009} showed, among other things, that the set of real numbers $\lambda > 0$ such that the ternary expansion of $\lfloor 2^n \lambda \rfloor$ avoids $2$ for infinitely many $n$ has Hausdorff dimension bounded below $1$.  See also de Faria and Tresser~\cite{defaria-tresser}, who conjecture that for any integer $b > 0$, in the base-$b$ expansion of the product of any $n$ primes that does not include all of $b$'s prime factors, the digits $\{0,1,\ldots,b-1\}$ become asymptotically equidistributed as $n \to \infty$.

Like our main conjecture, Conjecture~\ref{conj:two-thirds} involves the length of ``runs'' that an adversary can arrange.  Certainly he needs to use at least one of the first four coefficients before the terms become too small to sum to $1$.  However, if he can achieve a partial sum extremely close to $1$ in an initial subsequence, making the error term much smaller than $(2/3)^k$ in the first $k$ terms, then he can have a long string of zero coefficients before he needs to correct the sum further.  We believe that he cannot do this repeatedly, creating arbitrarily long runs of zeros with only a few nonzero coefficients between each run, but we have no proof.

\section{Numerical results}
\label{sec:numerics}

As mentioned earlier, the design we have proposed has the attractive computational property that a fractional-distance bound to depth $n$ can be verified by checking the minimum over just $\Theta(2^n)$,
 rather than $\Theta(4^n)$,
 distances between pairs of paths. Namely, if we write
\[
1-\delta_\ell 
= \max_{z \in \zset/2^\ell} \Re \left[ \frac{1}{\ell} \sum_{m=0}^{\ell-1} e_{2^\ell} \!\left( \left(\frac{2}{3}\right)^{\!m} z \right) \right] \, ,
\]
then our main conjecture is that $\sup_\ell (1-\delta_\ell) < 1$, and verification of the distance property for a tree of depth $n$ requires bounding the above ``$\max$'' for all $\ell \leq n$.
Applying Lemma~\ref{lem:inprod-to-hamming}, and specifically~\eqref{eq:kappa}, we then obtain an asymptotically good tree code with alphabet size $\kappa$ for 
\[
\kappa > \frac{2\pi}{\cos^{-1} \sup_\ell (1-\delta_\ell)} \, . 
\]

We can speed up the computation even further by employing a branch-and-bound strategy as follows.  Use a data structure which maintains at any time some of the nodes of the tree (initially just the root), and for each of these nodes, the real part of sum of $\beta$ along the the path from the root. At all times, maintain the record-holding lower bound for $1-\delta_\ell$ found so far (initially the record is $-1$). In each round, ``pop'' a node from the data structure; if it is possible for some descendant of the node to break the record, if the remainder of the path to that descendant turns out to consist solely of $1$'s, then ``push'' both children of the node onto the data structure.  Empirically, this strategy runs in time $O(2^{c\ell})$ where $c \approx 0.27$.

\begin{figure}
\begin{center}
\includegraphics[width=6.3in]{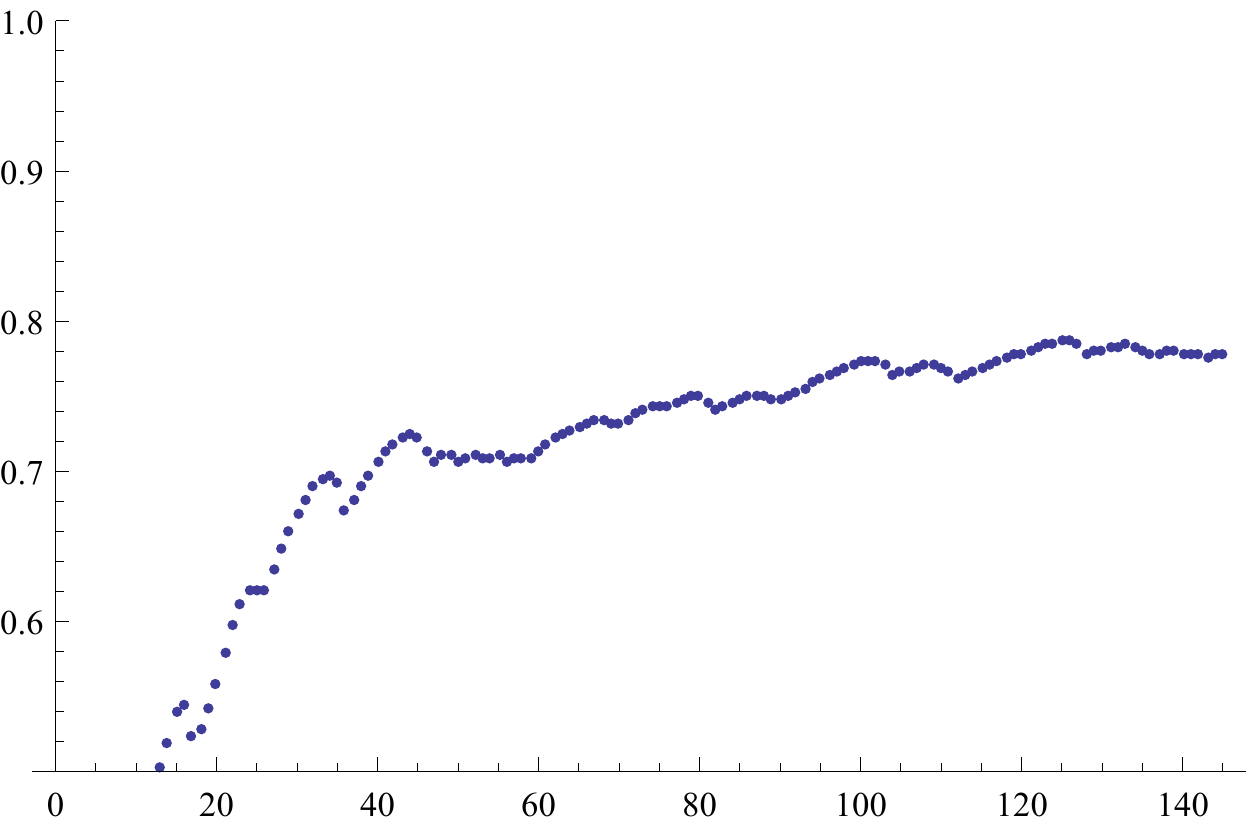}
\end{center}
\caption{$1-\delta_\ell$ plotted up to $\ell=145$; for the numerical values see Appendix~\ref{num_data}. ($1-\delta_\ell < 0.5$ for $\ell \le 12$.)}
\label{fig:DeltaTo145}
\end{figure}

We implemented this strategy in Mathematica (see Appendix~\ref{app:program}) and used it to compute $1-\delta_\ell$ up to $\ell=90$.  In this range, the largest $1-\delta_\ell$ is $1-\delta_{88}=0.7512$. Later, this problem was presented as a branch-and-bound exercise in a Caltech undergraduate algorithms course. Several teams of students, implementing various optimizations of the above basic strategy in C or C++, in some cases employing clusters of computers, were able to carry the computation even further.  
All of these calculations were conducted in double precision arithmetic, and the algorithm is numerically stable, so we have little concern about the accuracy aside from the possibility of programming errors. The latter risk is minimal due to the duplication of results by independently written programs; 
all results agreed to within $10^{-6}$ up to $\ell=99$, $10^{-4}$ up to $\ell=130$, and $2 \cdot 10^{-3}$ up to $\ell=145$.  
The team of Shival Dasu, Albert Gural, Nicholas Schiefer and Kevin Yuh computed $1-\delta_\ell$ up to $\ell=149$, the largest in this range being $1-\delta_{126}=0.7861$.
The team of Matt Mayers and Eric Wang computed $1-\delta_\ell$ up to $\ell=145$, and William Hoza's computation reached up to $\ell=133$. See Appendix~\ref{num_data} for the numerical values, which are also depicted in Figure~\ref{fig:DeltaTo145}.

If Conjecture~\ref{conj:main-conj} (equivalently~\ref{conj:main-conj-z}) should hold with $1-\delta = 0.9$, then we would obtain an asymptotically good tree code with alphabet size $\kappa=14$. 

We note incidentally that a simple greedy algorithm, that always chooses the square root of $\beta^3$ with positive real part, appears to achieve an asymptotic lower bound on $1-\delta_\ell$ of $ \approx 0.335$.
A somewhat better greedy algorithm, that takes the root whose argument is between $-\pi/3$ and $2\pi/3$, appears to converge to $1-\delta_\ell \approx 0.631$.  However, these are both much inferior to the true values found by the branch and bound algorithm.

\section*{Acknowledgments}

C.M. is supported by NSF grants CCF-1117426 and CCF-1219117, and ARO contract W911NF-04-R-0009. L.J.S. is supported by NSF grant CCF-1038578.  Many students' branch and bound implementations were very successful (23 students crossed the $\ell=100$ threshold); we thank all for their dedication and enthusiasm. Specific results are cited with permission. We also thank Alex Russell and Anup Rao for helpful discussions.

\bibliographystyle{plain}
\bibliography{refs}

\appendix
\section{Branch and Bound Program}
\label{app:program}

This program implements a branch-and-bound strategy for computing the maximum $1-\delta_\ell$ of the left-hand side of~\eqref{eq:main-conj-z} over all odd $z \in \zset/2^\ell$, or equivalently over all sequences where we start with $\beta = -1$ and, for $\ell-1$ steps, take one of the two square roots of $\beta^3$.  We prune nodes from the search tree whenever their subtree cannot match the largest real part we have seen so far, even if we assume that all subsequent terms contribute $1$.  When we go from $\ell$ to $\ell+1$, we simply assume that the next term has real part at least zero (since one of the square roots always does).

The main function \texttt{BandB[maxlen]} computes these bounds for all $2 \le \ell = \texttt{maxlen}$.  It returns an array \texttt{Relist} containing entries of the form $\{ \ell, N, z/2^\ell, 1-\delta_\ell \}$ where $N$ is the number of nodes explored, $z/2^\ell$ is the initial value of the worst sequence, and $1-\delta_\ell$ is the maximum achieved.

\medskip

\begin{verbatim}
(* branch-and-bound program for Conjecture 4 *)

(* t is an angle divided by 2pi *)
e[t_] := E^(2 Pi I t)
(* list of terms is stored in the form {sum, l, t} *)
score[terms_] := Re[terms[[1]]]

(* branch and bound: should we call off the search? *)
abandon[largrestResofar_, terms_] :=
  (score[terms] + len - terms[[2]] <= largrestResofar)

branch[terms_, lowerBound_] := Module[{nextt1, nextt2, tmp, t, sum, re, l},
  nodes++;
  sum = terms[[1]];
  re = score[terms];
  l = terms[[2]];
  t = terms[[3]];

  If[l == len,
   If[re > largestRe,
    largestRe = re;
    worstseq = t,
    ];
   Return[re]
   ];

  If[abandon[Max[largestRe, lowerBound], terms], Return[0],];

  (* next points to try: two square roots *)
  nextt1 := Mod[3 t/2, 1];
  nextt2 := Mod[(3 t + 1)/2, 1];
  If[1/4 < next1 < 3/4,
   tmp = next1;
   nextt1 = next2;
   nextt2 = tmp,
   ]; (* try root with positive real part first: saves some time *)

  branch[{sum + e[nextt1] // N, l + 1, nextt1}, lowerBound];
  branch[{sum + e[nextt2] // N, l + 1, nextt2}, lowerBound];
  Return[largestRe]
]

BandB[maxlen_] := Module[{lowerBound, Relist, worstlen, largestoverall},
  lowerBound = -1.;
  Relist = {};
  worstlen = 0;
  largestAvgRe = 0; (* worst among all l *)

  For[len = 2, len <= maxlen, len++,
   largestRe = -1;
   worstseq = {};
   nodes = 0; (* number of nodes searched *)
   (* initial list consisting of -1, -I *)

   branch[{-1 - I, 2, 3/4}, lowerBound];
   Print["l=", len,
     " nodes explored=", nodes,
     " largest average Re=", largestRe/len
   ];

   If[largestRe/len > largestAvgRe,
    largestAvgRe = largestRe/len;
    worstlen = len,
    ];

   lowerBound = largestRe;
   AppendTo[Relist, {len, nodes, worstseq, largestRe/len}];
   ];

  Print["worst overall: l=", worstlen,
   ", worst average Re=", largestAvgRe];
  Return[Relist]
]

(* check up to length 90 *)
BandB[90];
\end{verbatim}

\section{Numerical data for $\ell=1,\ldots,145$}
\label{num_data}
Computed values of $1-\delta_\ell$, in blocks of 10:

\scriptsize{
-1,
-0.5,
-0.09763107,
0.02244755,
0.06104173,
0.21035816,
0.30944832,
0.36931059,
0.38886774,
0.37897848,

0.43264236,
0.47419939,
0.50293127,
0.51902248,
0.53922673,
0.54395341,
0.52409779,
0.52756127,
0.54191323,
0.55916641,

0.57941523,
0.59694159,
0.61106656,
0.62006287,
0.62026151,
0.62094591,
0.63495312,
0.64792145,
0.65991213,
0.67092253,

0.68082984,
0.68926756,
0.6953662,
0.69724859,
0.69122805,
0.67276814,
0.68159584,
0.68993896,
0.69781049,
0.70519257,

0.71200453,
0.718033,
0.7227827,
0.72517324,
0.72299356,
0.71226017,
0.70661433,
0.70962287,
0.70963933,
0.70543927,

0.70942025,
0.71112421,
0.70837005,
0.70860827,
0.7097255,
0.7064789,
0.7078381,
0.70891003,
0.70956335,
0.71419406,

0.71841608,
0.72193815,
0.7241241,
0.72616951,
0.72938946,
0.7313194,
0.73375238,
0.73427986,
0.73113851,
0.7311115,

0.73415884,
0.73760456,
0.74065397,
0.74295896,
0.74377264,
0.74217703,
0.74522741,
0.74783505,
0.7495795,
0.74957126,

0.74606473,
0.74129791,
0.74394143,
0.745946,
0.74665782,
0.74908835,
0.75083691,
0.75119509,
0.74874461,
0.74762941,

0.7504024,
0.75311471,
0.75576782,
0.75836253,
0.7608983,
0.76337161,
0.76577249,
0.76807669,
0.77022843,
0.77210255,

0.77342368,
0.77360272,
0.77145276,
0.76491854,
0.76515789,
0.76717188,
0.76890072,
0.77005378,
0.77001304,
0.76844771,

0.76712548,
0.76278831,
0.76488747,
0.76694975,
0.76897601,
0.77096703,
0.77292332,
0.77484488,
0.77673065,
0.7785772,

0.78037586,
0.78210629,
0.78372238,
0.78512108,
0.78607595,
0.78610236,
0.78422349,
0.77874941,
0.77903028,
0.78064159,

0.78211905,
0.78333206,
0.78399579,
0.78351048,
0.78071408,
0.77733999,
0.77850478,
0.77909491,
0.779437,
0.77883906,

0.77796443,
0.77750376,
0.77602639,
0.7769396,
0.77717729
}
\end{document}